# Mobile Edge Computing

**Sohaib Ahmed**
FAST School of Computing
FAST NUCES Lahore
Lahore, Pakistan
l215635@lhr.nu.edu.pk

**Muhammad Hassan Khalid**
FAST School of Computing
FAST NUCES Lahore
Lahore, Pakistan
l215692@lhr.nu.edu.pk

**Muhammad Hamza**
FAST School of Computing
FAST NUCES Lahore
Lahore, Pakistan
l215636@lhr.nu.edu.pk

**Danyal Farhat**
FAST School of Computing
FAST NUCES Lahore
Lahore, Pakistan
danyal.farhat@lhr.nu.edu.pk

**Abstract**:

Mobile Edge Computing (MEC) has emerged as a solution to the high latency and suboptimal Quality of Experience (QoE) associated with Mobile Cloud Computing (MCC). By processing data near the source, MEC reduces the need to send information to distant data centers, resulting in faster response times and lower latency. This paper explores the differences between MEC and traditional cloud computing, emphasizing architecture, data flow, and resource allocation. Key technologies like Network Function Virtualization (NFV) and Software-Defined Networking (SDN) are discussed for their role in achieving scalability and flexibility. Additionally, security and privacy challenges are addressed, underscoring the need for robust frameworks. We conclude with an examination of various edge computing applications and suggest future research directions to enhance the effectiveness and adoption of MEC in the evolving technological landscape.

## Introduction:

Mobile Edge Computing (MEC) has emerged as a pivotal paradigm in telecommunications, revolutionizing the utilization of computational resources on the network edge. This study offers a comprehensive overview of MEC, focusing on its integration with existing and emerging technologies. MEC brings computational capabilities closer to end-users, facilitating low-latency, high-bandwidth applications and enabling new services such as augmented reality and IoT. However, despite its promise, MEC faces several challenges that must be addressed for its widespread adoption and optimization. Key areas of focus include scalable system architectures, energy-efficient computing, robust security mechanisms, resource management, and integration of advanced technologies for automated network management. By addressing these challenges, stakeholders can fully harness the potential of MEC, transform network infrastructure, and deliver enhanced user experiences.

## Literature Review:

1. **Scalable and Adaptive System Architectures**

   Wang et al. [4] emphasize the importance of efficient mobility management strategies to address frequent handovers between edge servers. There is a critical need for the development of MEC architectures that can dynamically adapt to changing network conditions and user demands [7]. This includes the creation of scalable system designs that can efficiently manage the increase in workload and support the dynamic nature of mobile applications and services [8].

2. **Energy-Efficient Computing**

   Innovations in mechanisms for more efficient energy transfer, along with advanced computational offloading strategies, are crucial to support sustainable MEC operations. Abd-Elnaby et al. [6] provide insights into novel trends and potential research areas, guiding future developments in MEC architectures, applications, and standards. The challenge lies in devising comprehensive system optimization techniques that balance energy consumption with computational requirements, especially in dynamic and resource-constrained environments.

3. **Unified Security Mechanisms**

   Security and privacy concerns are significant in the landscape of Mobile Edge Computing (MEC) deployment landscapes, necessitating robust measures to counter potential intrusions and data breaches (Abbas et al., [5]). Given the sensitive nature of data processing at the network edge, preserving data confidentiality, integrity, and authentication stands as paramount. Both Abbas et al. The discussion of unified security mechanisms underscores the urgency of addressing these challenges. A critical research gap lies in formulating cohesive security frameworks that are adept at seamlessly integrating MEC and its accompanying network infrastructure, including 5G and beyond. Such frameworks must combat a wide range of cyber threats using advanced encryption techniques, access controls, and intrusion-detection systems. Harmonizing these efforts is vital for safeguarding the user data and network infrastructure.

4. **Resource Management and Optimization**

   Efficient resource management is a cornerstone challenge in Mobile Edge Computing (MEC), as emphasized by Mao et al. [1]. This challenge encompasses the intricate tasks of balancing computational loads, optimizing resource utilization, and ensuring Quality of Service (QoS) amidst the dynamic nature of wireless environments [2]. Central to addressing this challenge is the strategic allocation of computational and communication resources, a process that is significantly influenced by informed decisions regarding computational offloading [10]. The decision-making process, as outlined by the proposed formula, considers various factors, including job memory requirements, available device memory, job priority, and estimated execution time on the device.

   $$O_{d,j} = \begin{cases} 1 & \text{if } M_j \leq M_d \text{ and } w_j \leq E_{d,j} \\ 0 & \text{otherwise} \end{cases}$$

   For instance, a job requiring 500MB of memory may be feasibly offloaded to a device with 750MB of available memory, exemplifying scenarios in which offloading proves to be advantageous. However, challenges have emerged, such as insufficient device memory and extended execution times for high-priority jobs, potentially diminishing the efficacy of offloading strategies. As such, navigating these complexities requires the dynamic adaptation of resource-allocation algorithms to maintain performance levels and meet application demands in the face of evolving network conditions.

5. **Seamless Integration with 5G Technologies**

   Integrating Mobile Edge Computing (MEC) with traditional cloud infrastructures poses technical challenges, particularly concerning workload distribution and resource allocation [9]. Zhang [3] underscores the complexity in optimizing task offloading decisions and resource allocation policies between MEC servers and centralized cloud data centers, aiming for efficient computational task balancing. To address this challenge, seamless integration frameworks and efficient communication protocols must be developed to ensure the interoperability and resource scalability in heterogeneous environments. Furthermore, as MEC continue to evolve, there is a pressing need for research on their effective amalgamation with 5G and forthcoming technologies. Comprehensive studies are warranted to fully exploit MEC's potential of MEC in enhancing network performance and user experience by integrating advanced 5G features such as NOMA, mmWave, and massive MIMO. This holistic approach is vital for realizing the synergistic benefits of MEC and 5G technologies for optimizing network operations and delivering enhanced services to users.

# Methodology:

**Edge Cloud Architecture:**

The Three-Tier Edge-Cloud Orchestration Architecture can be a suitable solution for scalable and adaptive architectures for mobile edge computing by distributing resources and tasks across multiple layers, reducing latency, and ensuring efficient resource utilization, making it suitable for handling large-scale IoT deployments and dynamic workload variations. It is a sophisticated system devised to manage a substantial volume of data and tasks originating from various Internet of Things (IoT) devices, such as smart home gadgets or sensors, by distributing these responsibilities across three distinct tiers:

1. IoT Device Tier: IoT devices, including new 5G-enabled and AI-powered devices, decide whether to execute tasks locally or delegate them to a more capable computing entity.

2. Edge Server Tier: More powerful than IoT devices but less powerful than cloud servers; these edge servers incorporate advanced hardware, such as GPU accelerators and FPGA chips. They handle tasks from IoT devices that require immediate responses and can offload less time-sensitive tasks to the cloud during high workloads.

3. Cloud Layer: Highly scalable cloud servers, including serverless and containerized architectures, handle tasks that can tolerate some latency, and support edge servers near capacity.

The architecture employs secure, low-latency 5G and Wi-Fi 6 communication protocols between IoT devices and edge servers, thereby minimizing unnecessary data exchange. Task placement is determined using a two-step methodology that leverages machine learning.

- **IoT Layer:** IoT devices assess their memory availability, task urgency, and energy consumption requirements to ascertain whether a task should be forwarded to edge servers.
- **Edge Layer:** Edge servers prioritize and promptly execute urgent tasks. However, they may delegate less urgent tasks to the cloud layer if they encounter excessive workload.

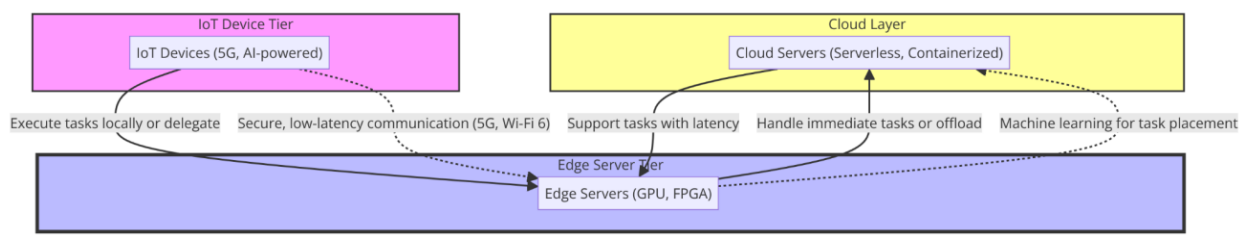

To ensure seamless operation, the system incorporates intelligent resource-allocation strategies at the edge level. This approach ensures that each task is allocated sufficient resources without wastage, thereby enabling the system to efficiently handle more tasks.

In essence, this architecture facilitates the intelligent and efficient handling of data and tasks from numerous devices, distributing the workload strategically to prevent bottlenecks and guarantee rapid responses, particularly for urgent tasks.

**Formulating the Offloading Problem**

The Delay-Aware Energy-Efficient (DAEE) [12] offloading algorithm is designed to save energy and meet task deadlines by using a virtual queue that mimics real tasks. This method minimizes the long-term energy use by considering deadlines and network conditions. By creating a virtual queue through Lyapunov optimization, the algorithm can make smart offloading decisions in real-time without missing deadlines. It sets up a system model with actual and virtual queues to efficiently handle tasks. The system model is represented by the following equation for the distance between the $ith$ device and the

MEC server: $d_i(t) = \sqrt{(x_i(t) - x_0(t))^2 + (y_i(t) - y_0(t))^2}$ where $(x_i(t), y_i(t))$ denotes the position of device i, and $((x_0(t), y_0(t))$ represents the MEC server's location.

**Lyapunov-Guided Optimization**

The DAEE [12] algorithm uses Lyapunov optimization to turn the offloading issue into a control problem to keep the virtual queue stable. The Lyapunov function is crucial in this process as it shows the overall congestion in the system, with actual and virtual queue backlogs represented by specific variables $Q_i(t)$ and $H_i(t)$ respectively.

$$L((t)) = \frac{1}{2N} \sum_{i=1}^{N} \left( Q_i(t)^2 + H_i(t)^2 \right)$$

The goal is to minimize a function to balance energy use and speed by making instant offloading choices based on the current system status without knowing future events. The offloading decisions are made in real-time, based on the system's current state, without requiring knowledge of future events.

**Hierarchical Federated Learning**

In Mobile Edge Computing (MEC), security is a major concern because of the distribution setup. To address this, robust security measures are required to protect data privacy and integrity in a changing 5G environment. The challenge is to create a single security system that can defend against various cyber threats, which is a key area of MEC research.

Hierarchical Federated Learning (HFL) can help fill this security gap. It works by improving the security in a decentralized environment through a dual strategy.

1. **Local Computation**: Defined by the equation $w_{ci}(k+1) = w_{ci}(k) - \eta \nabla F_i(w_{ci}(k))$
   This process enables individual edge devices to independently update their security parameters based on local data. This allows for the refinement of security measures at the device level, utilizing first-hand experiences with potential threats while maintaining data confidentiality.

2. **Cluster Aggregation**: This is encapsulated in the equation $w_c = \frac{\sum_{i=1}^{X} |D_{ci}| w_{ci}}{|D_c|}$, where the model updates at the cluster level are aggregated. It synthesizes updates from the individual devices to form a unified security model. By leveraging collective learning from various devices, the overall security framework is enhanced without compromising data privacy.

3. In a scenario where three devices choose a security protocol version based on their data, Device A switches from versions 1 to 2 after analyzing five data points. Device B always chooses version 3 and Device C prefers version 5 after processing two data points. The decision on the best security protocol version was made by considering each device's choice in relation to the amount of data analyzed.

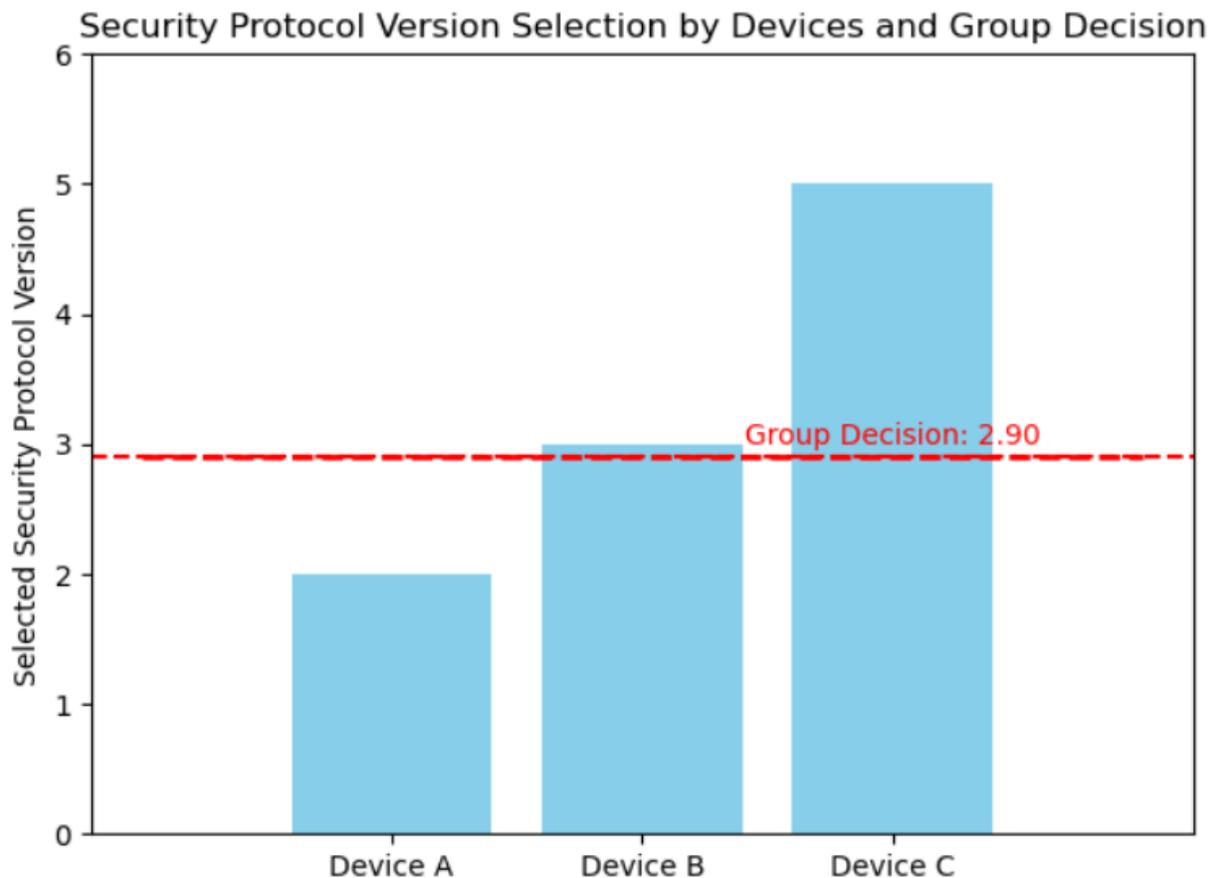

The bar chart shows how individual preferences lead to group decisions regarding the preferred security protocol version for the network. The HFL model, through its sophisticated integration of local computation and cluster aggregation, addresses the critical need for a unified security mechanism in MEC. It provides a scalable, flexible, and privacy-preserving approach that is well suited to the decentralized nature of MEC. Moreover, by enabling a collective defense strategy that evolves with each new data point and threat encountered, the security framework remains robust and adaptive to the complexities of future network technologies. This method not only fills the identified research gap but also aligns with the essential requirements for securing MEC environments in the era of 5G and beyond.

## Implementation

### Integration of NFV and SDN

Network Function Virtualization (NFV) and Software-Defined Networking (SDN) are transformative technologies that significantly contribute to solving resource management challenges in Mobile Edge Computing (MEC) environments. By leveraging these technologies, the MEC infrastructure can achieve unprecedented levels of flexibility, scalability, and efficiency in managing computational, storage, and networking resources.

The NFV decouples network functions from proprietary hardware, enabling them to operate as software applications on general-purpose servers. This shift allows for the agile deployment, scaling, and management of network services, transforming rigid physical networks into flexible, software-driven environments. NFV's ability to virtualize network services means that resources can be allocated dynamically based on current demands. For instance, functions such as load balancing, firewalls, and intrusion detection systems can be quickly instantiated as virtual network functions (VNFs) wherever needed to optimize resource utilization and reduce deployment times.

On the other hand, SDN introduces a centralized control mechanism over network devices, separating the control plane from the data plane. This separation allows network administrators to manage traffic routing and network policies through software independent of the underlying hardware. With SDN, network resources can be allocated and adjusted in real time through software controllers, based on the analysis of network conditions, application requirements, and predictive insights. This dynamic management capability ensures that the bandwidth is efficiently distributed, latency is minimized, and network configurations can be adapted to changing demands without manual intervention.

The integration of NFV and SDN in MEC environments addresses several resource-management challenges.

1. **Dynamic Resource Allocation**: By virtualizing network functions and centralizing network control, NFV and SDN enable dynamic allocation of computational and networking resources. Resources can be efficiently redistributed in response to fluctuating demand, thereby ensuring optimal performance and user satisfaction.
2. **Scalability and Flexibility**: The virtualization of network functions and centralized management of network policies allow MEC infrastructures to scale resources up or down as needed. This scalability and flexibility support the efficient handling of peak loads and rapid deployment of new services.
3. **Reduced Latency**: With ability to manage traffic flows and network functions through software, NFV and SDN can help minimize latency by optimizing routing decisions, prioritizing critical data, and placing services closer to the edge of the network.
4. **Energy Efficiency**: By optimizing resource allocation and reducing reliance on physical hardware, NFV and SDN contribute to energy savings. Virtualized network functions consume less power than traditional hardware-based solutions, and dynamic resource management ensures that resources are not wasted on the underutilized functions.
5. **Operational Cost Reduction**: The reduced need for specialized hardware combined with the efficiencies gained through dynamic resource management leads to significant cost savings. Operational expenses are reduced because network services can be deployed and managed with greater ease and fewer physical resources.

In conclusion, NFV and SDN have revolutionized resource management in MEC by introducing levels of agility, efficiency, and intelligence that were previously unattainable with traditional network architectures. By virtualizing network functions and centralizing control, these technologies enable MEC infrastructures to dynamically adjust to the demands of applications and users, optimize resource utilization, and enhance the overall performance and reliability of edge computing environments.

**DL and ML for Automated Network Management**

According to Miranda McClellan, Cristina Cervelló-Pastor, and Sebastià Sallent, managing thousands of heterogeneous connections under strict response constraints for applications, service creation, and network administration presents a complex challenge to 5G networks using edge computing. To realize the benefits of edge computing, it is necessary to develop automated procedures to provide, orchestrate, and manage network services and applications under conditions that change over time and across localities. A promising solution is to introduce machine learning (ML) to network operations to meet this new set of demands that are beyond the limitations of traditional optimization techniques.

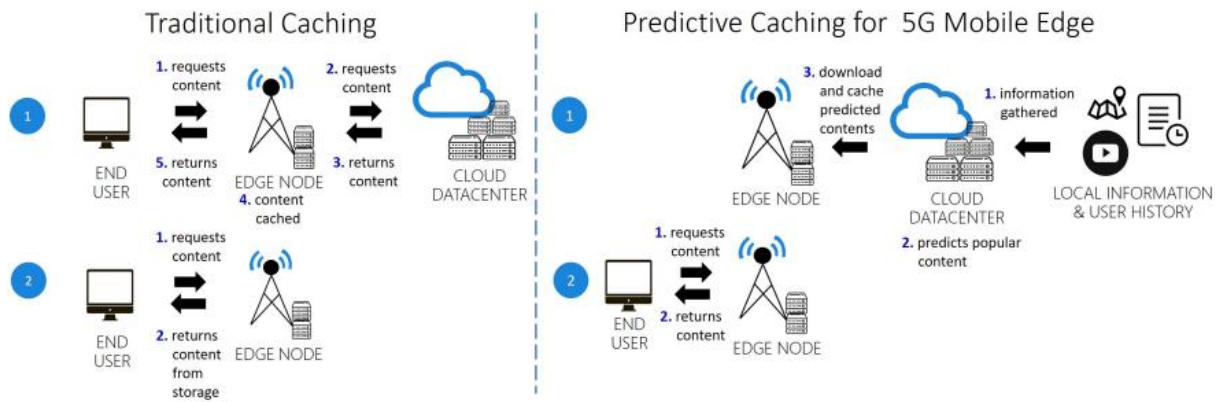

Figure. These two images show the difference between traditional caching and predictive caching using machine learning in mobile edge computing. In Step 1 of predictive caching, the most popular content that matches the user's predicted preferences according to their profile is downloaded from the cloud to an edge node. In the second step, when the user requests a specific content, there is a higher probability that the desired content has already been downloaded to the edge node previously, increasing QoE [11]

## Use Cases:

In the period spanning From 2021 to 2024, significant strides have been made in the realm of technology, particularly concerning the integration of 5G technologies within Mobile Edge Computing (MEC) frameworks. A notable development in this field is the application of Federated Learning to 5G Traffic Forecasting. This method employs federated learning algorithms to accurately predict network traffic patterns, thereby enabling network operators to make real-time adjustments to the network resources. Such predictive capabilities are instrumental in ensuring the efficient allocation of MEC resources, which in turn sustains high performance levels and minimizes latency for edge applications. Additionally, Artificial Intelligence (AI) has been pivotal in bridging the technological gaps in the 5G infrastructure. This is particularly evident in the automotive sector, where AI plays a crucial role in the development of autonomous vehicles. AI integration in mobile devices, such as Meta and Vision-Pro glasses, harmonizes with devices and services, creating a unified digital experience tailored to lifestyles.

## Results Obtained:

This section presents the dynamic distances between multiple mobile devices and a Mobile Edge Computing (MEC) server over time. The results are pivotal for understanding potential offloading opportunities, optimizing network resources, and enhancing the overall quality of service in edge computing environments.

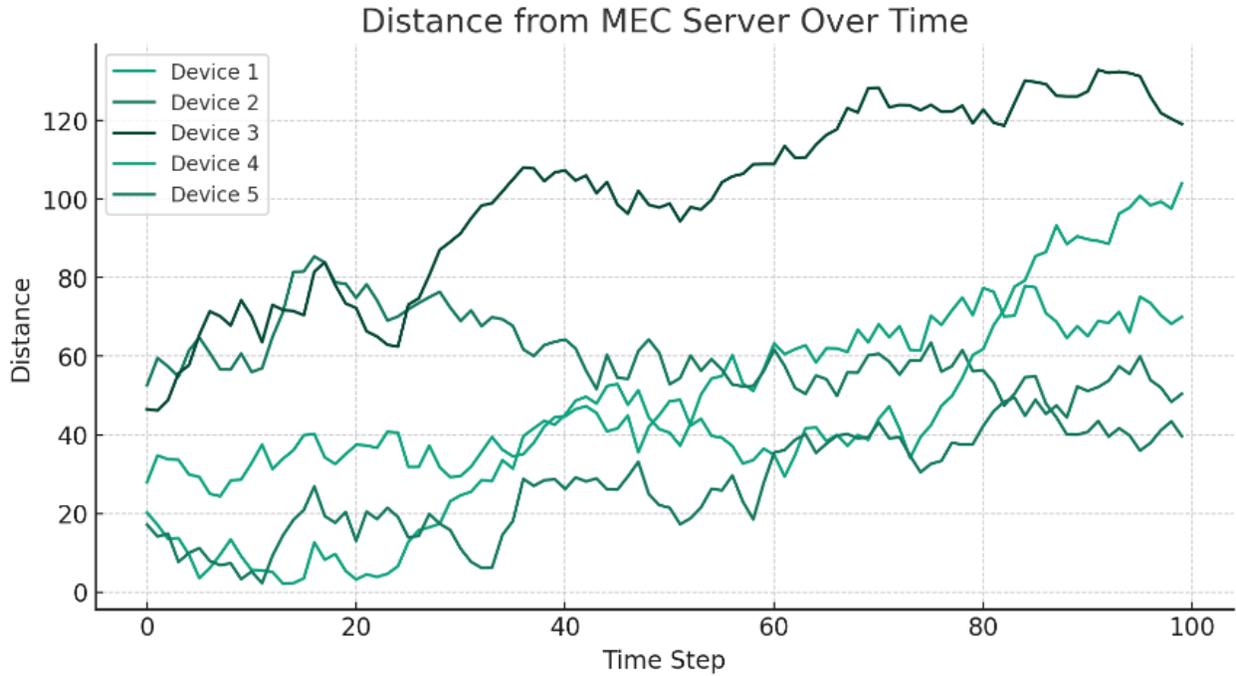

Figure: This plot illustrates the variation in distances between five different devices and the MEC server over time, highlighting the mobility and dynamic nature of edge-computing environments.

The plot below displays the fluctuating distances of the five devices from the MEC server as both devices move over time. Each line represents a different device, showing how their respective distances change at each time step. This visualization aids in understanding the spatial dynamics that are crucial for network management and decision-making in real-time applications.

The table provides a detailed view of the distances between each device and the MEC server over the first five-time steps, which is the resultant table for the above plot. Each cell shows the distance in meters, offering insight into the proximity of the devices to the server, which is essential for decision-making regarding data offloading and resource allocation.

| Time Step | Device 1 | Device 2 | Device 3 | Device 4 | Device 5 |
|---|---|---|---|---|---|
| 0 | 20.21 | 52.62 | 46.51 | 27.97 | 17.15 |
| 1 | 17.12 | 59.54 | 46.20 | 34.76 | 14.16 |
| 2 | 13.53 | 57.40 | 48.83 | 33.79 | 14.86 |
| 3 | 13.74 | 54.77 | 55.67 | 33.67 | 7.69 |
| 4 | 9.57 | 61.43 | 57.72 | 29.94 | 9.96 |

## Future directions:

Future directions for Mobile Edge Computing (MEC) encompass several key areas of research and development aimed at enhancing the capabilities and addressing the challenges of edge computing environments:

- Focus on developing dynamic orchestration mechanisms to efficiently handle the rapid expansion of edge resources while seamlessly integrating with existing network infrastructures.
- Explore innovative approaches to optimize energy consumption, such as integrating renewable energy sources, dynamic power management techniques, and designing energy-aware algorithms for task offloading and resource allocation.
- Develop robust security mechanisms to protect user data and infrastructure integrity, leveraging techniques such as homomorphic encryption, secure multi-party computation, and blockchain-based authentication.
- Apply artificial intelligence (AI) and machine learning (ML) algorithms for predictive resource allocation, workload scheduling, and network optimization, as well as investigate federated learning approaches for collaborative decision-making and adaptive resource provisioning.
- Explore novel architectures and protocols to enable interoperability and collaboration between heterogeneous edge platforms, fostering the development of innovative edge applications and services.
- Develop industry standards and open-source frameworks for seamless integration and interoperability across diverse edge environments, enabling developers to build and deploy applications across multiple platforms with ease.
- Prioritize the development of user-centric edge services that leverage context-awareness, personalization, and adaptive interfaces to deliver personalized experiences across various domains.

## Conclusion of Results:

These results underscore the variability in device-server proximity, which can significantly influence the efficiency of mobile edge computing applications. By meticulously analyzing these distances, network operators can make informed decisions that optimize resource allocation, enhance service quality, and reduce operational costs. This analysis is crucial for deploying efficient and responsive mobile edge computing frameworks, particularly in environments that require low latency and high data throughput.

## Conclusion:

In conclusion, this report comprehensively explores the landscape of Mobile Edge Computing (MEC), highlighting its pivotal role in revolutionizing telecommunications infrastructure. Through an in-depth analysis of scalable system architectures, energy-efficient computing strategies, unified security mechanisms, resource management, and seamless integration with 5G technologies, the review delineates critical challenges and proposes innovative solutions. Moreover, by presenting methodologies such as edge cloud architecture, formulation of offloading problems, integration of NFV and SDN, and leveraging DL and ML for automated network management, the study elucidates practical approaches to addressing these challenges. The examination of real-world use cases and the presentation of obtained results further underscore the significance of MEC in optimizing network operations and enhancing user experiences. Overall, this report underscores the imperative of addressing research gaps and technical challenges to fully realize the transformative potential of MEC in shaping future network infrastructures and advancing telecommunications capabilities.